\documentclass[twocolumn]{article}

\usepackage{smsec}
\usepackage{graphicx}
\usepackage{enumerate}

\Title{
Universality of Tsallis q-exponential of interoccurrence times \\ within the microscopic model of cunning agents
}

\Author{
Mateusz Denys$^1$,
Tomasz Gubiec$^1$,
and
Ryszard Kutner$^1$
}
\Affiliation{
$^{1}$Faculty of Physics, University of Warsaw, Ho\.za 69, PL-00681 Warsaw, Poland \\
}

\Email{mateusz.denys@fuw.edu.pl (correspondence author); tomasz.gubiec@fuw.edu.pl; ryszard.kutner@fuw.edu.pl;
}

\Abst{
We proposed the agent-based model of financial markets where agents (or traders) are represented by three-state spins located on the plane lattice or social network. The spin variable represents only the individual opinion (advice) that each trader gives to his nearest neighbors. In the model the agents can be considered as cunning. For instance, although agent having currently a maximal value of the spin advises his nearest neighbors to buy some stocks he, perfidiously, will sell some stocks in the next Monte Carlo step or will occupy a neutral position. In general, the trader has three possibilities: he can buy some stocks if his opinion change within a single time step is positive, sell some stocks if this change is negative, or remain inactive if his opinion is unchanged. The predictions of our model, found by simulations, well agree with the empirical universal distribution of interoccurrence times between daily losses below negative thresholds following the Tsallis q-exponential.}
\Keywords{
Agent-based model, Cunning agent, Return, Loss, Interocurrence time, Simulation, Tsallis q-exponential
}

\begin{document}
\maketitle


Agent-based modeling is a fruitful modern trend being yet a challenge not only for financial market description \cite{SZSL} but, e.g., for economical, social and environmental sciences \cite{BFPS} -- all of them deal with extraordinary complex systems.
Such a modeling constitutes a bridge between micro- and macroscales of system activities and enable to identify the laws that govern them. For instance, in the case of financial markets these laws lead to estimation of a risk investment \cite{IOC2}.

We restrict our considerations to study properties of financial markets. We merge two microscopic, agent-based socio-econophysical approaches explicitly taking into account opinions of agents: (i) the threshold model of the social impact type \cite{SH,BL,LNL,VN,HKS} with (ii) the concept of the negotiation round inspired by the Iori model \cite{Iori}.
We consider $N=1024$ interacting agents (or traders) on a square lattice of linear size $n=32$ (where $N=n\times n$) represented by three-state spin variable $s_j = 0, \pm 1$, $j = 1, \ldots, N$. The value of $s_j$ represents only an opinion or advice which $j$-th agent gives to his nearest neighbor in a single time step or a single spin drawing $t$.
Value $s_j = +1$ means the advice to buy stocks, while $s_j=-1$ to sell them. Value $s_j = 0$ simply means no advice or a neutral opinion of the agent. 
After each drawing, the chosen spin $s_i(t),\; i=1,2,\ldots,N,$ is updated according to the threshold linear social impact rule,
\begin{eqnarray}
s_i(t)=\mathrm{sgn}_{\lambda |M(t)|}\left[I_i(t)+\epsilon_i(t)\right],  
\label{s_i(t)}
\end{eqnarray}
where local social impact function
\begin{equation}
I_i(t)=\sum_{j=1}^{N}J_{ij}s_{j}(t)
\label{I_i(t)}
\end{equation}
and the threshold characteristic
\begin{equation}
\textrm{sgn}_{Y}(x)=\left\{ \begin{array}{lll}
-1 & \textrm{if} & x  < -Y,\\
0 & \textrm{if} & -Y \leq x < Y,\\
+1 & \textrm{if} & x \geq Y;
\end{array} \right.
\label{sign}
\end{equation}
the control parameter (threshold amplitude) $\lambda $ is a positive value. The strength of pair interaction $J_{ij}>0$ if agent $j$ is one of the four nearest neighbors of the agent $i$, otherwise $J_{ij}=0$. 
The (local) additive noise term, $\epsilon_i(t)$, represents the own, temporal, random opinion of agent $i$. In our simulation we used the additive noise distribution in the form of the Weierstrass-Mandelbrot probability density function (cf. Eq. (7) in Ref. \cite{DGK}) which can assume both L\'evy and non-L\'evy forms asymptotically. 
As usual, the temporal magnetization, $M(t)$, of the network, is defined as a mean of the current spin values.
Accordingly, the magnetization represents in this paper the aggregated opinion of traders. Apparently, quantity $\lambda |M(t)|$ in the definition of a single-state  characteristics (\ref{sign}) is a basic temporal threshold defining states of traders. Another (negative) threshold, $-Q$, defines the border between admissible and non-admissible losses. Both thresholds play a decisive role in our modeling.
Notably, rule (\ref{s_i(t)}) has a quite different meaning from all other ones used earlier in this context. That is, this rule considers the agent opinions instead of agent activities.

The change of any spin can affect its neighbors within a single round. The round, consisting of $N$ drawings of spin values, is a righteous time step as it gives \underline{in average} each trader a single chance to change his opinion. The single round is, in fact, 1MCS/spin used already in the dynamical Monte Carlo methods and is considered as a time unit, while a single drawing or a single time step is simply 1MCS.
This results from our basic local rule of the opinion dynamics which significantly differs from its all other counterparts used hitherto. The commonly used non-linear social impact rule, containing in (\ref{s_i(t)}) the product $s_iI_i$ instead of $I_i$, is improper in our case. This is because the non-linear rule leads to unrealistic situations, e.g., when given trader can change its opinion from the negative one to the positive opinion even if opinions of all his neighbors are negative and own random opinion of the agent vanishes.

The activity of the agent requires two subsequent single time steps leading to the change of the spin value, $d_i(t)=s_i(t)-s_i(t-1)$, i.e. to the change of the agent's opinion during subsequent drawings. For $d_i>0$ we deal with the agent demand, while for $d_i<0$ with the agent supply. The agent buy stocks if his spin value increases in the current step in comparison with its value in the previous step. The agent sell stocks if the value of his spin variable decreases. 


Usually \cite{RJB}, in agent-based models one considers the formula of a price formation, where logarithmic return, $S_{\tau }$, is proportional to the excess demand, $ED_{\tau}$. That is,
\begin{eqnarray}
S_{\tau }(t)=\ln{P(t)} - \ln{P(t-\tau)}=\frac{1}{\Lambda }ED_{\tau}(t)
\label{ror}
\end{eqnarray}
where
\begin{eqnarray}
ED_{\tau}(t)&=&\sum_{i=1}^{N}[s_{i}(t) - s_{i}(t-\tau )] \nonumber \\
&=&N[M(t) - M(t - \tau )],
\label{ror1}
\end{eqnarray}
$\tau $ (counted in rounds) is a delayed time and $\Lambda $ is a depth of the market.
Apparently, the excess demand, $ED_{\tau }(t)$, can change only if the opinion of any agent changes. Obviously, opinions' changes of various traders can mutually cancel making no influence on excess demand. Hence, price $P$ changes if and only if the mean opinion, $M$, changes, as it is required by a real-life market.



Now, we explain how a possible trapping (equivalent to the vanishing of market liquidity) is avoided in our simulation. By trapping we understand, herein,
a fully ferromagnetic state. Then the market has a great chance of being trapped for a long time by this extreme magnetization state. To avoid this trapping effect the system was activated by an exogenous factor, which can play the role of a market maker, performing an abrupt transition of the system to the paramagnetic 
state\footnote{The abrupt transitions observed from time to time are a characteristic feature of modern financial markets.}. 
Next, the evolution of the system is continued and the above given analysis of the system is repeated until subsequent abrupt transition.

\begin{figure}
\begin{center}
\includegraphics[width = 0.45\textwidth]{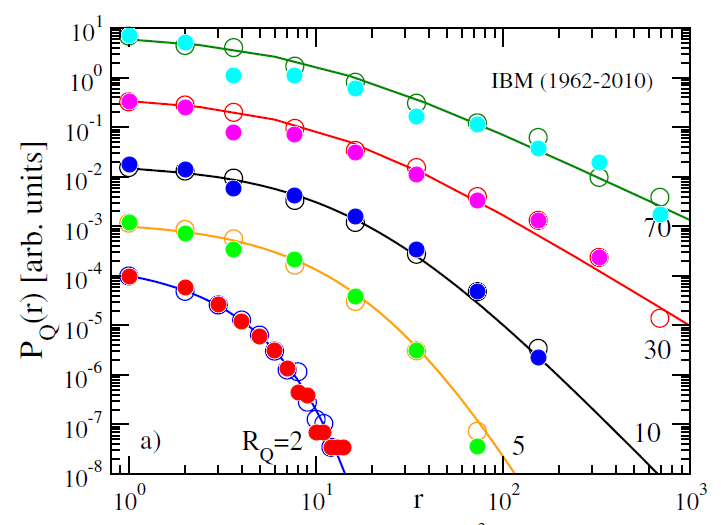} \\
\includegraphics[width = 0.24\textwidth]{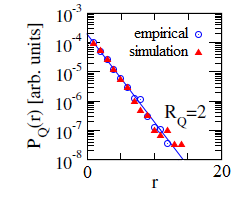}
\includegraphics[width = 0.19\textwidth]{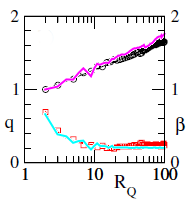}
\end{center}
\caption{Distribution function $P_Q(r)$ vs. interoccurrence time $r$ for daily price returns (see Ref. \cite{IOC2} for details), for instance, of IBM company in the period 1962-2010 (the upper and the lower-left plots -- our full plot for quotations of 16 significant indices and companies was shown in Fig. \ref{figure:Fig4_Prog}). The empirical data (empty circles and squares taken from Ref. \cite{IOC2}) belong to $R_Q=2,5,10,30,$ and $70$. Solid curves show the q-exponentials given by Eq. (\ref{q-exp}). Full circles, full triangles and both broken curves shown in the lower-right plot present for comparison the corresponding results of our simulations for delayed time $\tau =10^3\, [round]$ (equivalent to a single trading day) and for properly chosen control parameters: $J=1,\; \lambda =2,\; K=5,\; b=2,\; b_0=0.2$ (hence, Pareto exponent $\beta =\ln K/\ln b > 2$, cf. Eqs. (7) and (9) in Ref. \cite{DGK}). Notably, the additive noise was only used, herein, as no fluctuation of coupling parameter $J$ was necessary to fit the empirical data.
}
\label{figure:AO14}
\end{figure}
\begin{figure}
\begin{center}
\includegraphics[width = 0.45\textwidth]{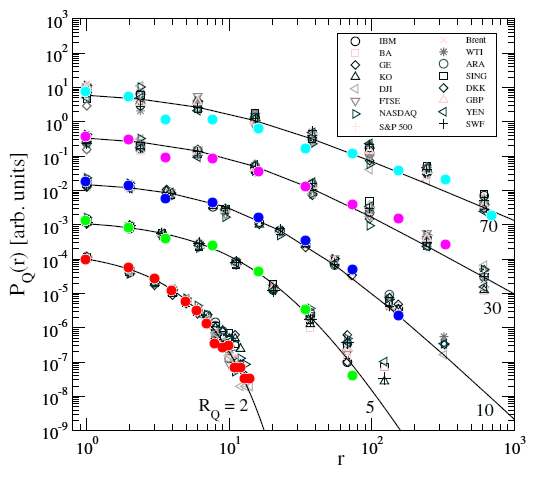}
\end{center}
\caption{Full plot of distribution function $P_Q(r)$ vs. interoccurrence time $r$ for the daily price returns of 16 quotations (significant market indices and companies) in the period 1962-2010. The empirical data (open marks, $\ast ,\; +$, and $\times $) taken from Ref. \cite{IOC2} are shown for $R_Q=2,5,10,30,$ and $70$. Solid curves show the q-exponentials given by Eq. (\ref{q-exp}). Full circles present the corresponding results of our simulations for fixed delayed time $\tau =10^3\, [round]$, and for control parameters $J=1,\; \lambda =2,\; K=5,\; b=2,\; b_0=0.2$. The additive noise was used, herein, as no fluctuation of coupling parameter $J$ was necessary to take into account.
}
\label{figure:Fig4_Prog}
\end{figure}

Main stylized facts coming from financial markets are already well reproduced by our model \cite{DGK}. Furthermore, we show in Figs. \ref{figure:AO14} and 
\ref{figure:Fig4_Prog} a comparison of our model predictions with recently discovered universal distribution of interoccurrence times in the form of the Tsallis q-exponential \cite{IOC2}. 

It has been shown that for returns, irrespective of the asset class, the distribution $P_Q(r)$ of the interoccurrence times between losses greater than some fixed negative threshold $-Q$ follows the Tsallis q-exponential form:
\begin{equation}
P_Q(r) \propto \frac{1}{\left[1+(q-1)\beta (R_Q)r\right]^{1/(q-1)}},
\label{q-exp}
\end{equation}
where parameter $q$ increases logarithmically with mean interoccurrence time $R_Q$ as follows \linebreak $q(R_Q) = 1 + q_0\ln(R_Q/2)$, where directional coefficient $q_0 
\approx 0.17$ (see Ref. \cite{IOC2} and the upper curves in the lower-right plot in Fig. \ref{figure:AO14} below). Coefficient $\beta (R_Q)$ decreases with $R_Q$ and above $R_Q \approx 15$ reaches a plateau having $\beta (R_Q>15)\approx 0.20$ -- see the lower broken solid curve in the lower-right plot in Fig. \ref{figure:AO14} (in Ref. \cite{IOC2} it was found $R_Q = 6$ and $\beta (R_Q>6)\approx 0.23$, respectively). 
One can see that our simulation results well agree with the empirical data -- this is the first description of the universality discovered in \cite{IOC2}, by the microscopic, agent-based model.

\end{document}